\documentclass[letterpaper, 10 pt, conference]{IEEEconf}  
\IEEEoverridecommandlockouts

\usepackage{cite}
\usepackage{amsmath,amssymb,amsfonts}
\usepackage{graphicx}
\usepackage{textcomp}
\usepackage{xcolor}
\def\BibTeX{{\rm B\kern-.05em{\sc i\kern-.025em b}\kern-.08em
    T\kern-.1667em\lower.7ex\hbox{E}\kern-.125emX}}

\usepackage{hyperref}
\usepackage{graphicx}		
\usepackage{wrapfig}
\usepackage[format=plain,font=footnotesize,labelfont=bf,labelsep=period]{caption}
\usepackage[export]{adjustbox}
\usepackage{bm}

\usepackage{amsmath}
\usepackage{mathrsfs}
\usepackage{amssymb}
\usepackage{mathtools}
\usepackage{algorithm}
\usepackage{algpseudocode}

\usepackage{pdfsync}
\usepackage[normalem]{ulem}
\usepackage{paralist}	
\usepackage[space]{grffile} 

\usepackage{color}

\usepackage{amsthm}
\usepackage{centernot}

\newtheorem{theorem}{Theorem}
\newtheorem*{theorem*}{Theorem}

\newtheorem*{lemma*}{Lemma}
\newtheorem{definition}{Definition}

\newtheorem*{proposition*}{Proposition}

\newcommand{\R}{\mathbb{R}}

\definecolor{darkblue}{RGB}{0,0,102}
\definecolor{lightblue}{RGB}{77,77,148}

\definecolor{gold}{RGB}{234, 170, 0}
\definecolor{metallic_gold}{RGB}{139, 111, 78}

\newcommand{\mb}[1]{\mathbf{ #1 }}
\newcommand{\bs}[1]{\boldsymbol{ #1 }}

\DeclareMathOperator*{\argmin}{argmin}

\newcommand{\lmat }{\begin{bmatrix}}
\newcommand{\rmat}{\end{bmatrix}}

\renewcommand{\P}{\mathbb{P}} 
\newcommand{\E}{\mathbb{E}} 
\usepackage{dsfont}

\newcommand{\codebase}{http://www.rkcosner.com/research/4-ICRA-GenerativeModeling/}

\usepackage{svg}

\begin{document}

\title{\LARGE \bf Generative Modeling of Residuals for Real-Time Risk-Sensitive Safety \\ with Discrete-Time Control Barrier Functions\\
}

\author{Ryan K. Cosner, Igor Sadalski, Jana K. Woo, Preston Culbertson, Aaron D. Ames
\thanks{ 
    This research is supported by BP.}
\thanks{The authors are with the Department of Mechanical and Civil Engineering, California Institute of Technology, Pasadena, CA 91125, USA. \texttt{$\{$rkcosner, isadalsk, jkwoo, pculbert, ames $\}$@caltech.edu}. 
}
}

\maketitle

\begin{abstract}
A key source of brittleness for robotic systems is the presence of model uncertainty and external disturbances. Most existing approaches to robust control either seek to bound the worst-case disturbance (which results in conservative behavior), or to learn a deterministic dynamics model (which is unable to capture uncertain dynamics or disturbances). This work proposes a different approach: training a state-conditioned generative model to represent the distribution of error residuals between the nominal dynamics and the actual system. In particular we introduce the Online Risk-Informed Optimization controller (ORIO), which uses Discrete-Time Control Barrier Functions, combined with a learned, generative disturbance model, to ensure the safety of the system up to some level of risk. We demonstrate our approach in both simulations and hardware, and show our method can learn a disturbance model that is accurate enough to enable risk-sensitive control of a quadrotor flying aggressively with an unmodelled slung load. We use a conditional variational autoencoder (CVAE) to learn a state-conditioned dynamics residual distribution, and find that the resulting probabilistic safety controller, which can be run at 100Hz on an embedded computer, exhibits less conservative behavior while retaining theoretical safety properties. 
\end{abstract}


\section{Introduction}


Robots operating in the real world face considerable uncertainty due to imperfect perception, approximate world and dynamics models, and random disturbances. These error sources are often a key failure cause for field-deployed robots, and in general can undermine classical safety and performance guarantees that rely on perfect models of the system and its environment. A natural framework to address this issue is risk-sensitivity: to operate effectively in novel, uncertain environments, robots should both be able to represent their uncertainty about the world (due to, e.g., modelling error, perception failures) and design controllers that can still (probabilistically) ensure some level of safety or performance, despite this uncertainty. 

In this paper, we explore using deep generative models (DGMs), \cite{rezende_variational_nodate, kingma_variational_nodate, kingma_auto-encoding_2022} which are a broad class of methods that use neural networks to approximate the probability distribution underlying a given dataset, to learn disturbance distributions for risk-sensitive control. 
These models can be used for density estimation, which provides a likelihood model for the data, and to sample new data points from the data distribution. Beyond their more traditional applications, like generating image \cite{dalle} and text \cite{gpt} data, these models have been applied to a broad range of robotics tasks including SLAM \cite{huang_creating_2022}, imitation learning \cite{urain_imitationflow_2020, chi_diffusion_2023}, motion planning \cite{carvalho_motion_nodate, ichter_learning_2018}, human-robot-interaction \cite{chi_collaborative_2020, xielearning}, anomaly detection \cite{park_multimodal_2018}, sim-to-real transfer \cite{inoue_transfer_nodate}, dynamics learning \cite{ren_learning_2020, watter2015embed},  and reinforcement learning \cite{hafner2019dreamer}.

\begin{figure}[t]
    \vspace{1em}
    \centering
    \includegraphics[width=\linewidth]{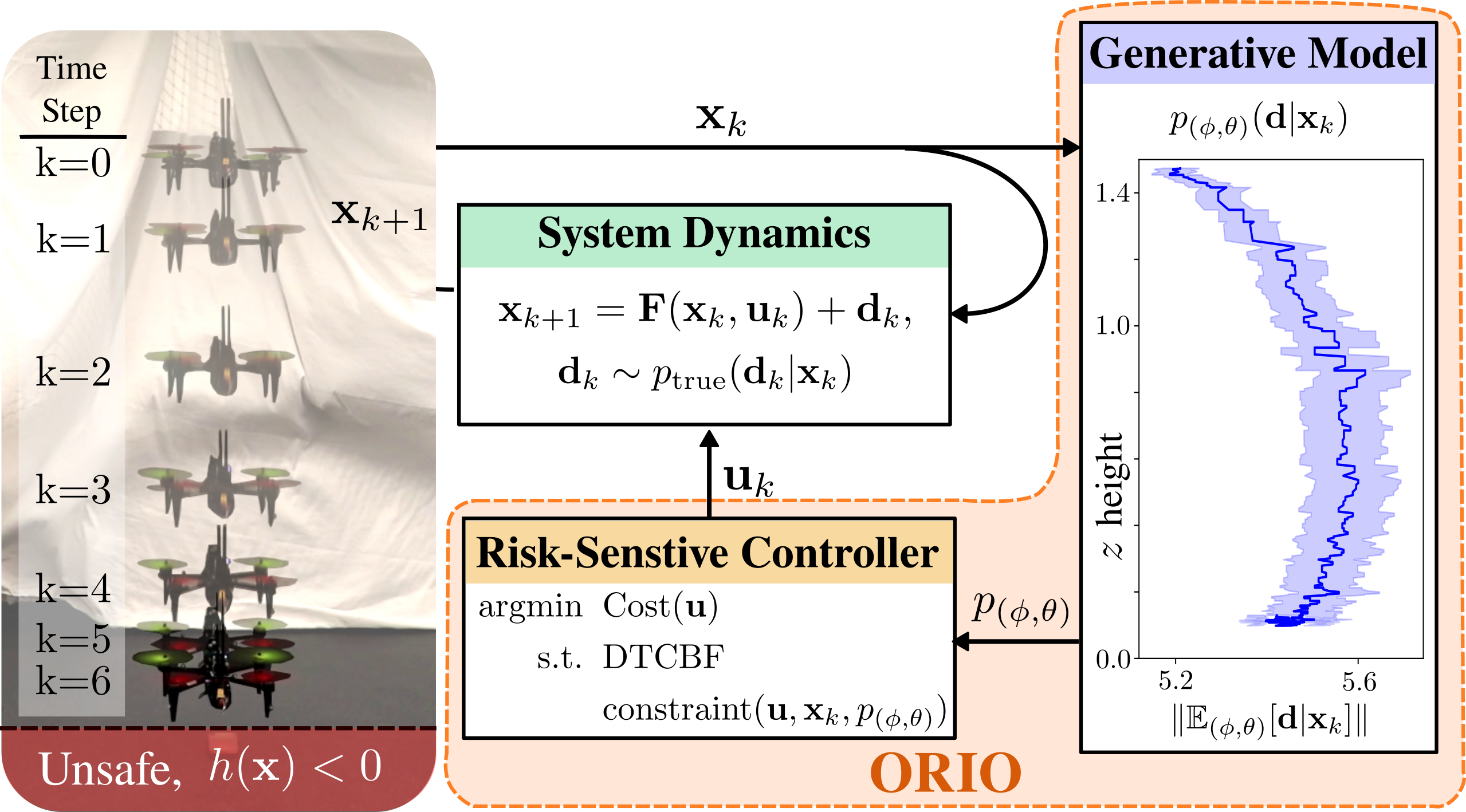}
    \caption{The drone falling through the air and avoiding hitting the ground using our proposed method. The generative model learns the state dependent disturbance distribution and the mean of the disturbance norm expanded by the trace of the covariance both scaled by the time step plotted on the right. }
    \label{fig:hero_figure}
    \vspace{-3.5em}
\end{figure}




In this work we employ conditional variational autoencoders (CVAEs) \cite{sohn_learning_nodate} which are a generalization of variational autoencoders (VAEs) that allow one to condition the generating process on a context variable (e.g., the current state). CVAEs have been used to recreate hand-written images of numbers given the desired digit \cite{sohn_learning_nodate} or to predict trajectories given state and environment understanding \cite{ivanovic_multimodal_2021,barbie_trajectory_2019,noseworthy_task-conditioned_nodate}. Since the generative process for a CVAE only requires two neural network forward passes and normal distribution samples, they are computationally efficient and well suited for real-time robotics applications. 


Stochastic control methods can make use of generative models to ensure constraint satisfaction (e.g., collision avoidance) up to a desired probability. In general, stochastic control methods assume prior knowledge of distribution such as value at risk (VaR) or conditional value at risk (CVaR) \cite{ahmadi_barrier_2020} values of a constraint function for the noisy dynamics, or the disturbance distribution's mean and covariance \cite{cosner_robust_2023,wei_moving_2022}. While this assumption is less restrictive than the global upper bound on the disturbance magnitude common in classical, deterministic robust control \cite{sontag_input--state_1995, cosner_measurement-robust_2021, bansal_hamilton-jacobi_2017}, it is still unrealistic to assume one has perfect, \textit{a priori} knowledge of the disturbance distribution before operating the system, and impractical / unprincipled to estimate these parameters by hand. To address these issues, we propose to learn a conditional generative model of the dynamics distribution instead of assuming its structure \textit{a priori}. 

This work on modeling dynamics residuals most closely resembles a probabilistic generalization of \cite{shi_neural_2019,oconnell_neural-fly_2022}. 
For generative modeling this work leverages  \cite{sohn_learning_nodate} and for safety we rely on the probabilistic safety guarantees for DTCBFs with stochastic dynamics residuals presented in \cite{cosner_robust_2023}. 

This work presents the Online Risk-Informed Optimization (ORIO) controller, a risk-based safety framework that learns to ensure safety in the presence of stochastic dynamics residuals using DGMs and DTCBFs. The main contributions of this work are (1) the proposed ORIO controller, which is a 
unified framework for dynamics distribution learning and usage of that distribution for ensuring probabilistic safety using DTCBFs; and (2) simulation and hardware demonstrations of the real-time application of these methods on a multi-rotor aerial robot for safe flight. 





\section{Conditional Variational Autoencoders (CVAEs) for Generative Disturbance Modeling}

In this work, we consider applications of safe control in the presence of unmodeled disturbances. Specifically, We consider the following systems with discrete time dynamics: 
\begin{align}
    \mb{x}_{k+1} & = \mb{F}(\mb{x}_k, \mb{u}_k) + \mb{d}_k, \quad \forall k \in \mathbb{N} \label{eq:ol_dyn}
\end{align}
with state $\mathbf{x}_k \in \R^n$, input $\mb{u}_k \in \R^m $, unmodeled residual dynamics $\mb{d}_k $ that take values in $\R^\ell$ and are sampled from some unknown distribution $p(\mb{d} | \mb{x})$, and modeled dynamics $\mathbf{F}: \R^n \times \R^m \to \R^n $. A state-feedback controller $\mb{k}: \R^n \to \R^m $ yields the discrete-time closed-loop system: 
\begin{align}
    \mb{x}_{k+1} = \mb{F}(\mb{x}_k, \mb{k}(\mb{x}_k)) + \mb{d}_k, \quad \forall k \in \mathbb{N}. \label{eq:cl_dyn}
\end{align}
The assumptions that the dynamics residuals are input-independent and unmatched is a general one often made in robust control theory for discrete-time systems \cite{rakovic2005set,marruedo2002input,lam2007robust,xie1993robust}.

\subsection{Conditional Variational Inference}

To account for the unmodeled disturbances, we first seek a generative model that can approximate the conditional distribution $p(\mb{d}|\mb{x})$ given a dataset $\mathcal{D} = \{ (\mb{x}_i, \mb{d}_i) \}_{i=1}^N $. We do this by fitting a parametric distribution to $\mathcal{D}$ which attempts to maximize the likelihood of the observed data with respect to the learned distribution. 


 While there exist many (learning- and learning-free) methods for generative modeling, in this paper, we look to Conditional Variational Autoencoders (CVAEs) \cite{sohn_learning_nodate}, a variant of Variational Autoencoders (VAEs) \cite{kingma_introduction_2019} that allows the learned models to be conditioned on observations, $\mb{x}$. CVAEs assume there exists a latent variable $\mb{z}$ which captures the ``unobserved'' information explaining any non-random variation in the data distribution $p$. For example, if the setting of robot safety, the latent codes $\mb{z}$ could represent state-dependent modeling errors, or other hidden variables (e.g., higher-order dynamics, time delays) that could influence the difference between the observed next state $\mb{x}_{k+1}$, and the modeled dynamics $\mb{F}(\mb{x}_k, \mb{u}_k)$.

Specifically, CVAEs represent the conditional distributions $p_{\theta}(\mb{d}|\mb{x}, \mb{z})$ and $q_\varphi(\mb{z} | \mb{x}, \mb{d})$, and the latent prior $p_\phi(\mb{d}| \mb{x})$ as multilayer perceptions (MLPs) with corressponding parameters $\theta, \phi, \varphi$, and seek to optimize these parameters such that the data likelihood, $p_{\theta, \phi}(\mb{d})$, marginalized over all states $\mb{x}$ and latent codes $\mb{z}$, is maximized. Traditionally $q_\varphi$ is called an ``encoder,'' since it maps states $\mb{x}$ and disturbances $\mb{d}$ to distributions over the latent codes $\mb{z}$, and similarly $p_\theta$ is called a ``decoder.'' While maximizing $p_{\theta, \phi}(\mb{d})$ exactly is intractable (and, since we do not have access to this distribution, as we are hoping to estimate it from data), we instead optimize the evidence lower bound (ELBO):
%
%
\begin{align}
    \log p_{\theta, \phi}  (\mb{d} | \mb{x}) \geq  \E_{q_{\varphi}} &[\log p_\theta (\mb{d} | \mb{x}, \mb{z})]\\  &- KL\big(q_\varphi(\mb{z} | \mb{x}, \mb{d}) \Vert p_\phi (\mb{z} | \mb{x})\big) \nonumber 
\end{align}
where $KL$ is the Kullback-Liebler divergence. In practice, each network represents its corresponding distribution as a conditional Gaussian, with, e.g. $p_\theta(\mb{d}| \mb{x}, \mb{z}) = \mathcal{N}(\mb{d}; \bs{\mu}_\theta(\mb{x}, \mb{z}), \bs{\Sigma}_\theta(\mb{x}, \mb{z})),$ where $\mathcal{N}(\cdot\; ;\mu, \Sigma)$ is the probability density function of a multivariate Gaussian with mean $\bs{\mu}$ and covariance $\bs{\Sigma}$, and $\bs{\mu}_\varphi, \bs{\Sigma}_\varphi$ are neural network outputs representing the sufficient statistics of this distribution. 


\begin{figure}
    \centering
    \includegraphics[width=0.8\linewidth]{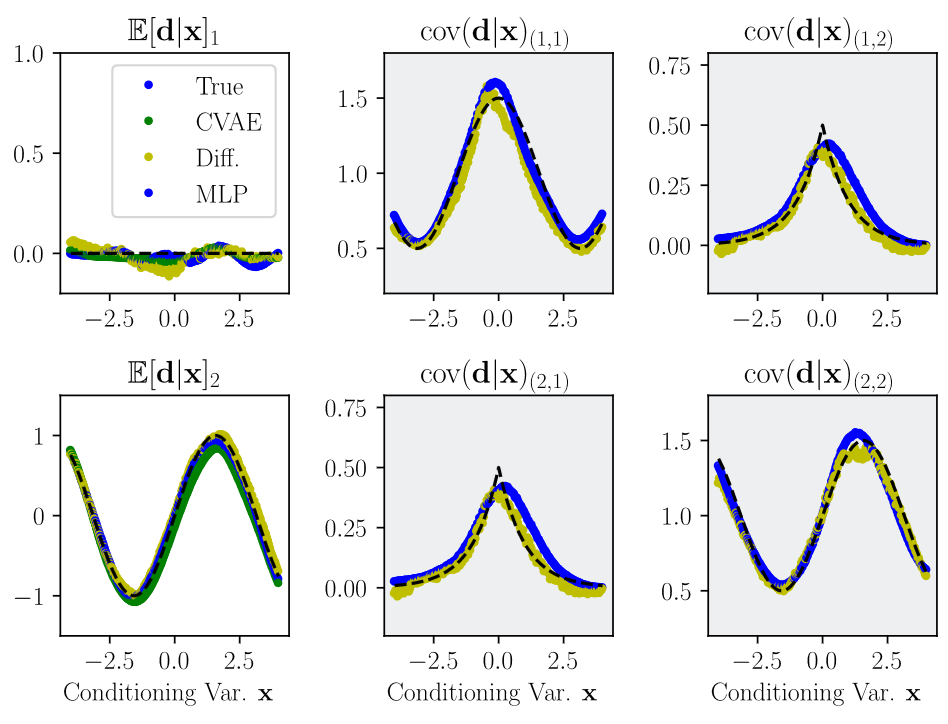}
    \caption{Learning heteroschedastic disturbance of double integrator system using 3 minutes of data at 100 Hz (36 five second long trajectories). The approximated mean and covariance values are scaled by the time step and plotted against the true values in black using the CVAE in blue, diffusion model in yellow, and MLP (mean only) in green.}
    \label{fig:heteroschedastic}
    \vspace{-3em}
\end{figure}

\subsection{Mean and Covariance Estimation using CVAEs}
Once trained, a CVAE can be used for estimation of the disturbance's conditional likelihood, generate new samples, or estimate the mean and covariance of the true distribution.

In particular, the risk-sensitive DTCBF-based controller \cite{cosner_robust_2023} we use requires us to compute the mean and covariance of the disturbance distribution $\mb{d}$. To do this, we use the following estimator\footnote{This estimator is best understood as a ``Rao-Blackwellization'' \cite{gelfand1990samplingbased} (as used in the Markov Chain-Monte Carlo literature) of the simple two-step sampling estimator which uses the sample mean and covariance of $\mb{d}^{(s)} \sim p_\theta(\cdot \mid \mb{x}, \mb{z}^{(s)})$. This estimator is known to have stronger convergence ($O(\frac{1}{\sqrt{S}})$) than the two-step scheme.} for $p_{\theta, \phi}(\mb{d} | \mb{x}):$
\begin{align}
    p_{\theta, \phi} (\mb{d}|\mb{x}) &  \approx \frac{1}{S}\sum^S_{s=1} p_{\theta}(\mb{d}|\mb{x}, \mb{z}^{(s)})  \label{eq:cvae_mc}\\
    & = \frac{1}{S} \sum_{s=1}^S \mathcal{N}\left(\mb{d} \; ; \; \mu_\theta(\mb{x}, \mb{z}^{(s)}), \Sigma_\theta(\mb{x}, \mb{z}^{(s)}) \right) \nonumber  
\end{align}
where $\mb{z}^{(s)} \sim p_\theta(\mb{z}|\mb{x}) $. Since this MC approximation is a Gaussian mixture model (GMM) we can obtain its mean and expectation in closed form as:
\begin{align}
    \E [\mb{d} | \mb{x} ]   \approx &    \frac{1}{S}\sum_{s=1}^S \mu_\theta(\mb{x},\mb{z}^{(s)})\label{eq:gmm_mean} \triangleq \overline{\bs{\mu}}(\mb{x}), \\
    \textup{cov}(\mb{d} |\mb{x})  \approx 
     &  \frac{1}{S} \Big(\sum_{s=1}^S  \Sigma_\theta(\mb{x}, \mb{z}^{(s)}) + \mu_\theta(\mb{x},\mb{z}^{(s)})\mu_\theta(\mb{x},\mb{z}^{(s)})^T \Big) \nonumber \\
      &\qquad -  \overline{\bs{\mu}}(\mb{x}) \overline{\bs{\mu}}(\mb{x})^\top \triangleq  \overline{\bs{\Sigma}} (\mathbf{x}) \label{eq:gmm_cov}
\end{align}

\subsection{Residual Dynamics Modeling: Simple Simulation}

\begin{figure}
    \centering
    \includegraphics[width=0.9\linewidth]{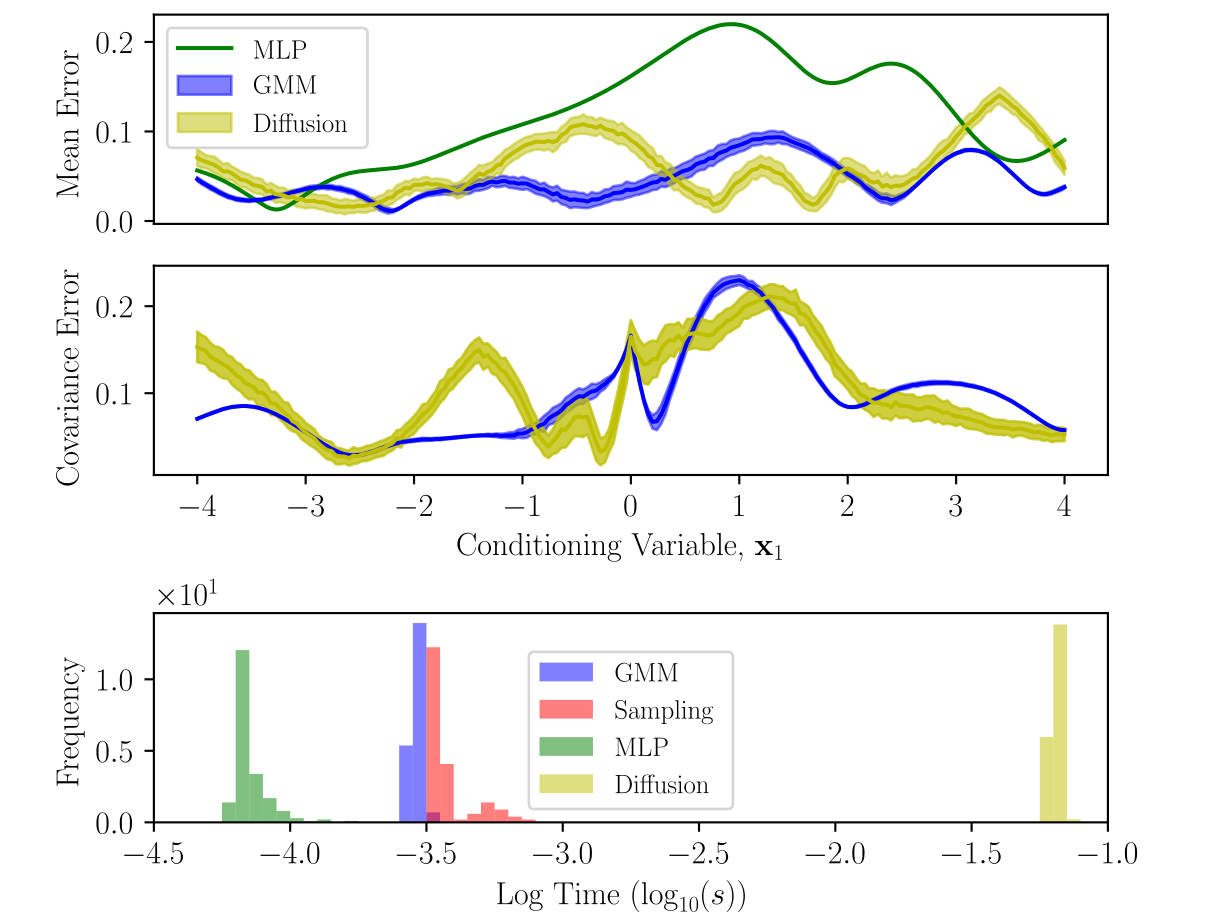}
    \caption{\textbf{(Top)} Mean error $\Vert \E_{\textup{predicted}}[\mb{d} | \mathbf{x}] - \E_{\textup{true}} [\mb{d} | \mathbf{x}]  \Vert$ vs. state for GMM-based CVAE sampling method (blue), MLP (green), diffusion model sampling mean estimate (yellow). One standard deviation estimated from 100 samples per state, $\mb{x}$ ,is plotted around the CVAE and diffusion model curves. \textbf{(Middle)} Covariance error $\Vert \textup{cov}_{\textup{predicted}}(\mb{d} | \mathbf{x}) - \textup{cov}_{\textup{true}} (\mb{d} | \mathbf{x})  \Vert_2$ vs. state for GMM-based CVAE method (blue) and diffusion model sampling-mean estimate (yellow) both using 10,000 samples is shown with one standard deviation. \textbf{(Bottom)} Evaluation time for each model to make its apprpoximation. Calculations performed on a desktop computer with a Nvidia 3090Ti GPU with 10,000 samples for the stochastic methods. }
    \label{fig:gmm_comparison}
    \vspace{-3em}
\end{figure}

To demonstrate the capabilities of CVAEs to learn complex dynamics disturbances, we consider the simple double integrator system: 
\begin{align}
    \mb{x}_k = \lmat x \\ v \rmat_{k+1} = \lmat 1 & \Delta_t\\
    0 & 1 
    \rmat \lmat x \\ v \rmat_k  + \mb{d}_k \label{eq:toy_dynamics}
\end{align}
with a state dependent residual distribution$, \quad \mb{d}_k \sim p(\mb{d}_k | \mb{x}_k) $ which is a state-dependent Gaussian distribution with mean $\bs{\mu}(\mb{x}) = \lmat 0,  &  \sin(x) \rmat^\top $ and the covariance is $\bs{\Sigma}(\mb{x}) = \frac{1}{2}\lmat 2 + \cos(x) & \exp(-|x|) \\ \exp(-|x|) & 2 + \sin(x)\rmat $. The system was initialized at $\mb{x}_0 = \mb{0}$, with $\Delta_t = 0.01$, and simulated for 35 five-second trials to collect data. Then the CVAE was trained to approximate the distribution. The CVAE accurately learns the nonlinear heteroschedastic disturbance with a relatively small amount (3 minutes, total, in simulation time) of data, as can be seen in Fig.  \ref{fig:heteroschedastic},


We compare the CVAE to two baselines: a conditional diffusion model \cite{ho2020denoising}, which is a state-of-the-art generative model that has recently seen interest as a policy representation for robotics \cite{chi_diffusion_2023}, and a simple MLP trained to map the state $\mb{x}$ to a fixed, deterministic disturbance $\mb{d}(\mb{x})$\footnote{Code for this test and the MLP and diffusion models can be found \href{\codebase}{here}.}. The results of this are shown in Fig.  \ref{fig:gmm_comparison} and Table \ref{tab:estimate_variance}. There we can see that the MLP, despite being significantly faster, tends to overfit to the noise causing higher mean error. Alternatively, the diffusion model accurately learns the distribution, but is nearly two orders of magnitude slower than the CVAE. Additionally, two approximation methods are used for the CVAE: the GMM-based estimator in (\ref{eq:gmm_mean}, \ref{eq:gmm_cov}) and a simple two-step sampling estimator using the population mean and covariance calculations from samples of $p_{\theta, \phi}(\mb{d}|\mb{x}) $. The GMM-based method is shown to be slightly faster and results in less average error and variance.

\begin{table}[]
    \centering
    \begin{tabular}{|c|c c|c c|}
        \hline & $\mu$ Err. Avg. $\pm 2 \sigma$ 
        & $\Sigma$  Err. Avg. $\pm 2 \sigma $\\
         \hline GMM  & 0.04512 $\pm$  0.00433  &  0.09518  $\pm$ 0.00296\\
         \hline Sampling & 0.04604 $\pm $ 0.00989 & 0.09710 $\pm$  0.01419\\
        \hline Diffusion & 0.05866 $\pm$ 0.00942 & 0.1025 $\pm$  0.01363\\ 
        \hline 
    \end{tabular}
    \caption{The statistics for the mean and covariance estimates of each estimation method obtained from 100 estimates at 201 states. The average error is similar for each model, but the GMM-based method has smaller variance which is important when using its outputs in closed-loop control. Estimates for each method are calculated using 10,000 samples. }
    \label{tab:estimate_variance}
    \vspace{-3em}
\end{table}



\section{Safety Theory using Discrete-Time Control Barrier Functions (DTCBFs)}

Now, given a way to learn complex, heteroscedastic noise distributions from trajectory data, we need a way to perform risk-sensitive control under this uncertainty. We begin by first formalizing our definition of safety. We define the safety of a system to be the forward invariance of some user-defined ``safe set'', $\mathcal{C}$, as is common in robotics and control theory \cite{brunke_safe_2021,ames_control_2019,wabersich_predictive_nodate,herbert_fastrack_2017}

\begin{definition}[Forward Invariance and Safety]
    A set $\mathcal{C}\subset\R^n$ is forward invariant for the system \eqref{eq:cl_dyn} if $\mathbf{x}_0 \in \mathcal{C}$ implies that $\mathbf{x}_k \in \mathcal{C}$ for all $k \in \mathbb{N}$. In this case, we call system \eqref{eq:cl_dyn} safe with respect to the set $\mathcal{C}$. 
\end{definition}

\subsection{Deterministic Safety with DTCBFs}

Discrete-time control barrier functions (DTCBFs) are tools for guaranteeing the safety of discrete time systems. Consider a set $\mathcal{C} \triangleq \{ \mb{x} \in \R^n ~:~ h(\mb{x}) \geq 0 \} $ expressed as the 0-superlevel set of a continuous function $h: \R^n \to \R$. 

\begin{definition}[Discrete-Time Control Barrier Function (DTCBF) \cite{agrawal_discrete_2017}]
    Let $\mathcal{C} \subset \R^n $ be the 0-superlevel set of a continuous function $h: \R^n \to \R$. The function $h$ is a discrete-time control barrier function (DTCBF) for \eqref{eq:ol_dyn} on $\mathcal{C}$ if there exists an $\alpha \in [0,1]$ such that for each $\mb{x} \in \R^n$, there exists a $\mb{u} \in \R^m $ such that: 
    \begin{align}
        h(\mb{F}(\mb{x}, \mb{u})) \geq \alpha h(\mb{x}). \label{eq:dtcbf} 
    \end{align}
\end{definition}

Given a CBF $h$ for \eqref{eq:ol_dyn} and a corresponding $\alpha \in [0,1]$, we define the point-wise set of control values: 
\begin{align}
    \mathscr{K}_\textup{CBF}(\mb{x}) = \left\{ \mb{u} \in \R^m : h(\mb{F}(\mb{x}, \mb{u})) \geq \alpha h(\mb{x}) \right\}
\end{align}
This yields the following theoretical result: 

\begin{theorem}(\cite{agrawal_discrete_2017})
    Let $\mathcal{C}\subset \R^n $ be the 0-superlevel set of a continuous function $h: \R^n \to \R $. If $h$ is a DTCBF for \eqref{eq:ol_dyn} on $\mathcal{C}$, then the set $\mathscr{K}_\textup{CBF}(\mb{x}) $ is non-empty for all $\mb{x} \in \R^n $, and for any continuous state-feedback controller $\mb{k}$ with $\mb{k}(\mb{x}) \in \mathscr{K}_\textup{CBF}(\mb{x})$ for all $\mb{x} \in \R^n $, then the system: 
    \begin{align}
        \mb{x}_{k+1} = \mb{F}(\mb{x}_k, \mb{k}(\mb{x}_k)) 
    \end{align}
    is safe with respect to $\mathcal{C}$. 
\end{theorem}
Intuitively, the value of $h(\mb{x}_k)$ can only decay as fast as the geometric sequence $\alpha^kh(\mb{x}_0) $, which is lower-bounded by 0, thus ensuring the safety (i.e., forward invariance) of $\mathcal{C}$. This inequality mimics that of a discrete-time Lyapunov function \cite{bof_lyapunov_2018} and similarly regulates the evolution of $h$ based on its previous value. 

Given a continuous nominal controller $\mb{k}_\textup{nom}: \R^n \times \mathbb{N} \to \R^m $ and a DTCBF $h$ for \eqref{eq:ol_dyn}
 on $\mathcal{C}$, a controller $\mb{k}$ satisfying $\mb{k}(\mb{x}, k)  \in \mathscr{K}_\textup{CBF}(\mb{x}) $ for all $\mb{x} \in \R^n $ and $k \in \mathbb{N} $ can be specified via the following optimization problem (assuming feasibility for all $\mb{x} \in \mathcal{C}$): 
 \begin{align}
     \mb{k}(\mb{x}) = \argmin_{\mb{u} \in \R^m } & \quad \Vert u -  \mb{k}_\textup{nom}(\mb{x}, k) \Vert^2 \\
     \textup{s.t.} & \quad h(\mb{F}(\mb{x}, \mb{u})) \geq \alpha h(\mb{x}) \nonumber 
 \end{align}

We note that unlike the continuous-time CBF constraint\cite{ames_control_2019}, the DTCBF inquality constraint \eqref{eq:dtcbf} is not necessarily convex with respect to the input, preventing it from being integrated into a convex optimization-based controller. To solve this issue, it is often assumed that the function $h\circ \mb{F} : \R^n \times \R^m \to \R$ is concave with respect to its second argument \cite{agrawal_discrete_2017,ahmadi_barrier_2020,zeng_safety-critical_2021}. This assumption was shown to be well motivated for concave $h$ and sufficiently fast sampling times \cite{taylor_safety_2022}. 

\subsection{Stochastic Safety with DTCBFs}
 
For this work we are interested in generalizing beyond deterministic DTCBFs by ensuring safety in the presence of unbounded disturbances, where the disturbance $\mb{d}_k$ is sampled from some probability distribution which may be a function of the state ($\mb{d}_k \sim p(\mb{d}_k | \mb{x}_k)$). 

Since system \eqref{eq:cl_dyn} will almost surely leave any compact set as time goes to infinity \cite{steinhardt_finite-time_2012,culbertson_input--state_2023}, we focus on bounding safety probabilities for a finite time horizon. 

\begin{definition}[$K$-Step Exit Probability \cite{cosner_robust_2023}]
    Let $h: \R^N \to \R$ be a continuous function. For any $K \in \mathbb{N} $, and initial condition $\mathbf{x}_0 \in \R^n$, the $K$-step exit probability of the closed-loop system $(10) $ is given by: 
    \begin{align}
        P_u(K, \mathbf{x}_0) = \P \left\{ \min_{0 \leq k \leq K} h(\mb{x}_k) < 0 \right\} 
    \end{align}
\end{definition}
\noindent In particular, this describes the probability that the system will leave $\mathcal{C}$ within $K$ steps. We now present a key result that bounds this exit probability when a CBF condition is imposed in expectation. 

\begin{theorem}[Thm. 5 \cite{cosner_robust_2023}] \label{thm:kushner}
   Let $h:\R^n \to \R$ be a continuous, upper-bounded function with upper bound $M\in\R_{>0}$. If there exists an $\alpha\in (0,1)$ such that the closed-loop system \eqref{eq:cl_dyn} satisfies: 
    \begin{align}
        \mathbb{E}[~h(\mb{F}(\mb{x}, \mb{k}(\mb{x})) + \mb{d})  \mid \mb{x}~] \geq \alpha h(\mb{x}),  \label{eq:kushner_constraint}
    \end{align}
    for all $\mb{x}\in \R^n$, with $\mb{d}\sim p(\mb{d} | \mb{x}) $, then for any $K \in \mathbb{N}$: 
    \begin{align}
    \label{eq:problo}
        P_u(K, \mb{x}_0) \leq 1 - \frac{h(\mb{x}_0) }{M}\alpha^K. 
    \end{align}
\end{theorem}

Since this constraint contains the expectation of a nonlinear function of the disturbance distribution, it may be difficult to compute. To this end \cite{cosner_robust_2023} provides control methods for ensuring that inequality is satisfied for systems where the disturbance mean and covariance are known:
\begin{theorem}[Thm. 6 \cite{cosner_robust_2023}]\label{thm:kushner_jensen}
Consider the system \eqref{eq:cl_dyn} and let $h:\R^n \to \R$ be a twice-continuously differentiable, concave function such that $\sup_{\mb{x} \in \R^n} h(\mb{x}) \leq M$ for $M\in\R_{>0}$ and  $\sup_{\mb{x} \in \R^n} \Vert \nabla^2 h(\mb{x}) \Vert_2 \leq \lambda_{\max} $ for $\lambda_{\max}\in\R_{\geq0}$. Suppose there exists an $\alpha \in (0,1)$  such that: 
\begin{align}
    \hspace{-0.5em} h(\mb{F}(\mb{x}, \mb{k}(\mb{x})) + \mathbb{E}[\mb{d} | \mb{x}] ) - \frac{\lambda_\textup{max}}{2}\textup{tr(cov}(\mb{d} | \mb{x})) \geq \alpha h(\mb{x}) \label{eq:jensen_dtcbf} 
\end{align}
for all $\mb{x} \in \mathcal{C}$ with $\mb{d} \sim p(\mb{d} | \mb{x})  $. Then we have that:
\begin{equation}
\mathbb{E}[~h(\mb{F}(\mb{x}, \mb{k}(\mb{x})) + \mb{d})  \mid \mb{x}~] \geq \alpha h(\mb{x}),
\end{equation}
for all $\mb{x}\in\mathcal{C}$ with $\mb{d}\sim p(\mb{d}|\mb{x})$. 
\end{theorem}

\noindent This provides a tractable way for enforcing safety once a model of the conditional mean $\E [\mb{d} | \mb{x}]$ and covariance $\textup{cov}(\mb{d}|\mb{x}) $ is known. 

In particular we propose the ORIO (Online Risk-Informed Optimization) controller: 
\begin{align}
     \mb{k}(\mb{x}) =\argmin_{\mb{u} \in \R^m } & \; \Vert u -  \mb{k}_\textup{nom}(\mb{x}, k) \Vert^2 \label{eq:orio} \tag{ORIO}\\
     \textup{s.t.} & \; h(\mb{F}(\mb{x}, \mb{u}) + \overline{\bs{\mu}}(\mathbf{x})) \nonumber \\ &  \quad \quad \quad \quad \quad -  \frac{\lambda_\textup{max}}{2}\textup{tr}(\overline{\bs\Sigma}(\mathbf{x}) )\geq \alpha h(\mb{x}) \nonumber 
 \end{align}
where we approximate $\E [\mb{d}|\mb{x}] $ and $\textup{cov}(\mb{d}|\mb{x}) $ using the outputs, $\overline{\bs{\mu}}_{(\phi, \theta)}(\mathbf{x}) $ and $\overline{\bs{\Sigma}}_{(\phi, \theta)}(\mathbf{x}) $, of the CVAE and use those approximations in conjunction with the simplified DTCBF constraint \eqref{eq:jensen_dtcbf} which endows a system with the $K$-step exit probability guaranteed by Thm \ref{thm:kushner}.

\begin{figure}
    \centering
    \includegraphics[width=0.8\linewidth]{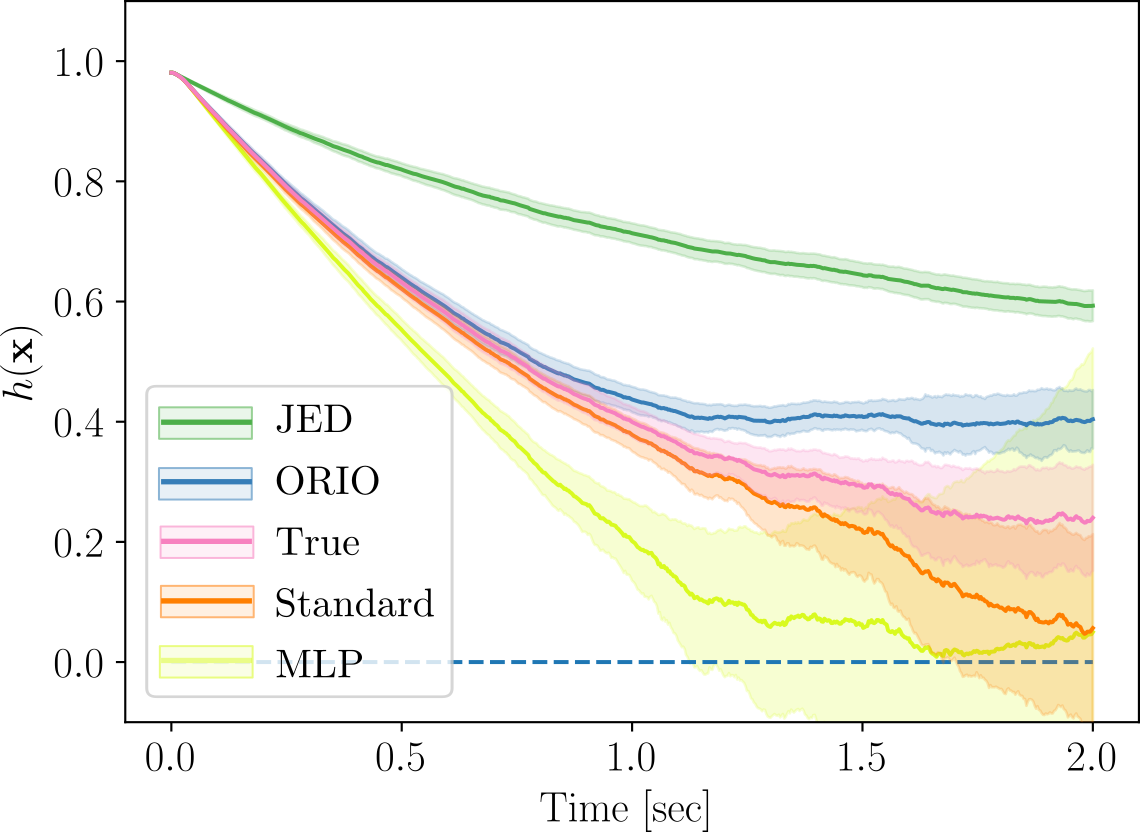}
    \begin{tabular}{|c|c|c|c|c|c|}
         \hline  $\P_\textrm{bound}$ & MLP & Standard & JED & True & ORIO \\
         \hline  0.82 & 0.69 & 0.56 & 0.00 &  0.32 & 0.11  \\ 
         \hline 
    \end{tabular}
    \caption{Quadrotor simulation Results. (\textbf{Figure}) The mean of 100 trajectories for each controller is plotted with 1/2 standard deviation around it. (\textbf{Table}) The $K-$step probability bound for the 2 second long trial from Thm. \ref{thm:kushner_jensen} and the approximated $K-$step probability experienced on in simulation over 100 trials.}
    \label{tab:my_label}
    \label{fig:enter-label}
    \vspace{-2em}
\end{figure}

\section{Safety for Quadrotor Drone: Theory and Experiments}

In this section we show the utility of our method which combines the usefulness of CVAEs as tools for learning complex dynamics residual distributions and our controller \eqref{eq:orio} which leverages those generative models for safe control. This framework is applied to safe flight on a quadrotor drone, and evaluated in both simulation and hardware.

\subsection{Modeling and DTCBF Synthesis for a Multirotor Drone}

We consider the safety of a quadrotor drone. We model the dynamics of this system as: 
\begin{align}
    \underbrace{\frac{d}{dt} \lmat \mb{p} \\ \mb{q} \\ \mb{v} \rmat}_{\dot{\mb{x}}} & = \lmat \mb{v} \\  \mb{0} \\ - \mb{e}_z g   \rmat + \lmat \mb{0} \\  \bs{\omega} \\ \frac{1}{m} \mb{R}(\mb{q}) \mb{e}_z  \tau \rmat 
\end{align}
where the state $ \mb{x} = (\mb{p} \in \R^3, \mb{q} \in \textup{SO}(3), \mb{v} \in \R^3) $ represents the position, orientation, and velocity, $g$ is gravity, $m$ is the drone mass, and the system has inputs of angular rate $\bs{\omega} \in \mathfrak{so}(3) $ and thrust force $\tau \in \R$. Here $\mb{e}_z$ is a unit vector in the $z$ direction and $\mb{R}: \textup{SO}(3) \to \R^{3 \times 3} $ maps the quaternion representation of orientation to the respective rotation matrix. For simulation,  these dynamics are approximated in discrete time using Euler integration on manifolds and for the dynamics approximation in the DTCBF, standard Euler integration is used to ease computation, \cite{tayal_control_2023} shows that this approximation is theoretically well justified for DTCBFs with short time steps.


The safety criteria for our quadrotor is to avoid collisions with the ground or roof. We can encode this safety as the 0-superlevel set of the function: 
\begin{align}
    h_{\textup{des}}(\mb{x}) = C - \bs{\zeta}^\top \mb{P} \bs{\zeta}  
\end{align}
for some $ C> 0$ where $\bs{\zeta} = \lmat z - z_0, & v_z \rmat $ and  $V(\bs{\zeta}) = \bs{\zeta}^T \mb{P} \bs{\zeta} $ is a Lyapunov function generated by the Discrete-time Algebraic Ricatti Equation (DARE) for discrete-time double integrator dynamics. 
However, this is not necessarily a DTCBF since the quadrotor's orientation may render it unable to track double integrator trajectories. 

To avoid this problem, we add an penalty term to ensure correct orientation when $h(\mb{x}) = 0 $. 
\begin{align}
    h(\mb{x}) = (C - \bs{\zeta}^\top \mb{P} \bs{\zeta} )  - \lambda ( 1 - \mb{e}_z R(\mb{q}) \mb{e}_z) \label{eq:drone_cbf}
\end{align}
 This DTCBF is motivated by the differential flatness of the multi-rotor dynamics \cite{mellinger2011minimum} since the system can track linear system trajectories. 
 
 Importantly, this is a valid DTCBF for the Euler-approximated dynamics and there are bounds for $\overline{\bs{\mu}} $ and $\frac{\lambda_\textup{max}}{2}\textup{tr}(\overline{\bs{\Sigma}}(\mathbf{x})) $ such that  \ref{eq:orio} controller is feasible for all $\mathbf{x} $ such that $h(\mb{x}) \geq 0 $. This is stated formally in the following theorem:

\begin{theorem} \label{thm:drone_safety}
    Consider $h$ as in \eqref{eq:drone_cbf} for $\alpha >0$. If $C \geq 2 \lambda$, then there exists $\mb{u}  \in \R^4$ and $M_{\bs{\delta}}, M_{\bs{c}}>0$ such that 
    \begin{align}
        h(\mb{F}_{\textup{Eul}}(\mb{x}, \mb{u}) + \bs{\delta})  + c \geq \alpha h(\mb{x}) 
    \end{align} for all $  \bs{\delta} < M_\delta,  c < M_c $ and all $\mb{x}$ such that $h(\mb{x}) \geq 0 $. 
\end{theorem}
\noindent See the appendix in section \ref{appendix} for the proof of this theorem. 




\subsection{Comparison Controllers}
For comparison, we implement several controllers in addition to \ref{eq:orio}. Each controller has the structure: 
 \begin{align}
     \mb{k}(\mb{x}) = \argmin_{\mb{u} \in \R^m } & \quad \Vert u -  \mb{k}_\textup{Nom}(\mb{x}, k) \Vert^2 \label{eq:comparison_controllers} \\
     \textup{s.t.} & \quad h(\mb{F}_\textup{Eul}(\mb{x}, \mb{u}) + \mathbf{m}(\mb{x})) - c(\mathbf{x})  \geq \alpha h(\mb{x}) \nonumber 
 \end{align}
 with the following ablations:
 \begin{itemize}
     \item \textbf{(Standard)} where $\mb{m}(\mb{x})  = 0 $ and $c(\mb{x})  = 0$. This is the standard DTCBF controller \cite{agrawal_discrete_2017} where the modeled dynamics are assumed to be correct.
     \item \textbf{(JED)} where $\mb{m}(\mb{x})$ is the constant sample mean of the $\mathcal{D}$ and $c(\mb{x}) $ is the trace of the sample covariance times $\sup_{\mb{x} \in \R^n} \Vert \nabla^2 h(\mb{x}) \Vert_2$. This is the ``Jensen-Enhanced DTCBF'' as presented in \cite{cosner_robust_2023}. 
     \item \textbf{(MLP)} where $\mb{m}(\mb{x})$ is an MLP that is trained on the dataset $\mathcal{D}$ to approximate the dynamics residuals and $c(\mb{x}) = 0 $.  
     \item \textbf{(True)} where $\mb{m}(\mb{x}) $ is the true dynamics residual mean and $c(\mb{x})$ is the trace of the true covariance times  $\sup_{\mb{x} \in \R^n} \Vert \nabla^2 h(\mb{x}) \Vert_2$.  
 \end{itemize}





\subsection{Simulation Results}

For simulation we use the dynamics residual model: 
\begin{align}
    p(\mb{d} | \mb{x}) = \mathcal{N}\bigg(\mb{d}; \underbrace{\mb{0}_{9}}_{\bs{\mu}(\mb{x}) }, \underbrace{I_9 \times (1 + 50e^{-30 z^2 } ) \times 10^{-5}}_{\textup{cov}(\mb{x})}\bigg) 
\end{align}
where the disturbance grows as the drone approaches the ground to approximate complicated ground effects.

We fly the drone using an SE(3) stabilization controller \cite{lee_geometric_2010} from 1 meter in the air to the ground 20 times for 2 seconds each with a control and data collection frequency of 333Hz to collect training data (13320 data points). Each controller was simulated for 100 two-second long trajectories at 333Hz with $\alpha = 0.9975$. Results for these simulations are shown in Fig. \ref{fig:heteroschedastic}. 
The looseness of the probability bound is in part due to the fact that the covariance is small for a large portion of the trajectory, which is not accounted for by the martingale-based bound.  Despite the risk probability bound not being tight \eqref{eq:orio}, produces behavior which is similar to (True) and which is less conservative than (JED) while still being more robust than (Standard) and (MLP).

\begin{figure}
    \centering
    \includegraphics[width=0.9\linewidth]{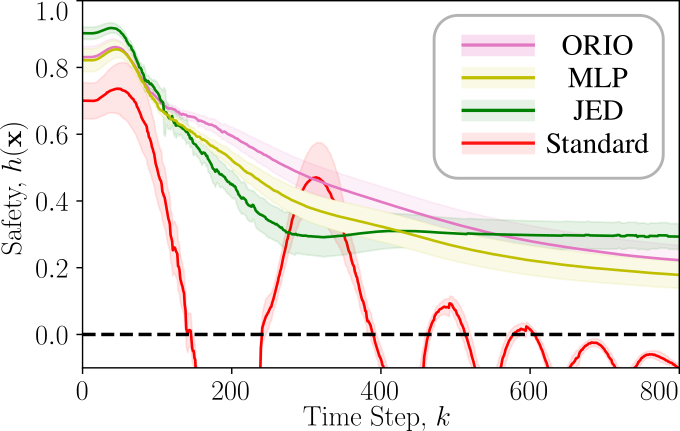}
    \caption{Mean and one standard deviation of $h(\mb{x})$ for the ``drop'' test case in hardware, which drops the drone with $\mb{k}_\text{nom}=0$ from a height of roughly two meters.  We compare our proposed controller \eqref{eq:orio}  with three ablations: a deterministic MLP, a simple aggregate across all trajectory data (JED), and a standard CBF controller. All controllers except the standard CBF satisfy the safety constraint $h(\mb{x}) \geq 0$, where the standard CBF fails due to the unmodelled dynamics. While the residual dynamics include complex aerodynamic effects, in this case they are low-variance, and so we expect both the MLP and \eqref{eq:orio} to perform similarly.}
    \vspace{-1em}
    \label{fig:drop}
\end{figure}


\subsection{Hardware Result}
Finally, we deploy our risk-informed controller \eqref{eq:orio} on a quadrotor drone flying aggressively near the ground. For all tests, we provide the drone with real-time pose measurements from a motion capture system. The drone is equipped with a Nvidia Tx2 that is used to perform onboard computation of all neural network forward passes and optimization-based controllers. The mean and covariance of the dynamics residuals are approximated using the CVAE with 200 samples at 100 Hz and the optimization problem in \ref{eq:orio} is an SOCP which is solved using en embedded conic solver \cite{domahidi2013ecos} at 300 Hz. Approximately 2 minutes of training data is collected via human-operated flight for both experiments. 

Our first experiment is a drop test (shown in Figure \ref{fig:hero_figure}) where we drop the drone from a hover at approximately 2m, and enforce the barrier constraint \eqref{eq:drone_cbf} with $\alpha = 0.9975$ for positions  above the ground; this case has low noise but requires accurate estimation of the quadrotor's thrust / ground effects for the barrier to be effective in preventing ground collision. Figure \ref{fig:drop} plots the mean barrier value $h(\mb{x})$ over at least 50 trials for each ablation of our method, with one standard deviation shaded around the mean. All controllers except the standard CBF (which nearly immediately becomes unsafe due to the unmodeled dynamics) exhibit conservative behavior and settle relatively far from the boundary. Of particular interest is the extremely similar behavior of the simple MLP and CVAE methods; this result is intuitive since the low-variance disturbance source allows the MLP to accurately capture the unmodeled dynamics. This provides an interesting insight: learning residual dynamics via simple regression, as proposed in \cite{zeng_tossingbot_2020,shi_neural_2019,oconnell_neural-fly_2022}, is well-posed for systems subject to deterministic, low-variance disturbances, and can yield safe, performant behavior without reasoning about stochasticity.

\begin{figure}
    \centering
    \includegraphics[width=0.8\linewidth]{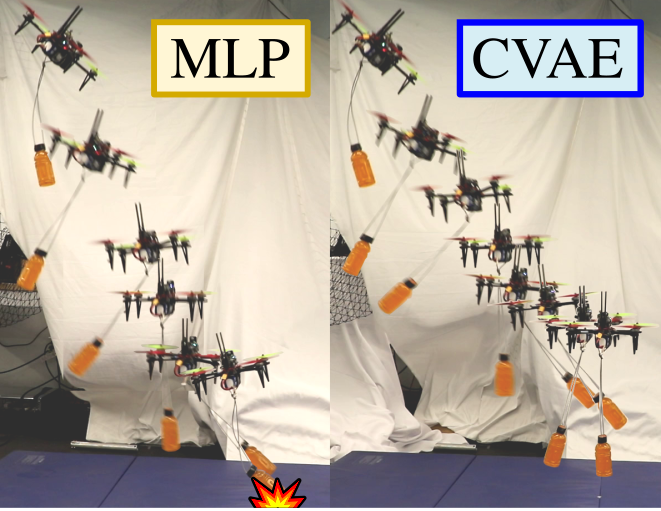}
    \includegraphics[width=0.9\linewidth]{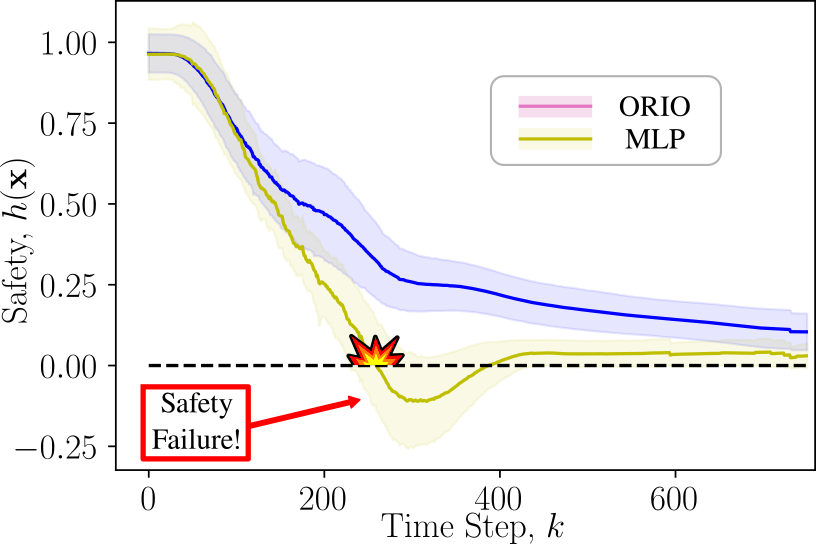}
    \caption{ \textbf{(Top)} The drone is dropped from the top left and moves to the right as it falls while carrying an orange payload. The left shows a failure case when controlled by the MLP controller and the right shows a success from the same initial condition when controlled by the ORIO controller. \textbf{(Bottom)} The average of 14 trajectories is plotted with one standard deviation shading. The ORIO controller successfully keeps the system safe while the MLP-based controller results in safety failures. The video of these experiments can be found \href{\codebase}{here}. }
    \vspace{-3em}
    
    \label{fig:hanging-mass}
\end{figure}

In our second test, the quadrotor is carrying a slung, unmodeled load of 0.55kg, which induces large disturbances that are not uniquely determined by the current state $\mb{x};$ this test requires the residual dynamics models to capture high-variance behavior to accurately model the slung load's effect on the dynamics. Here we again enforce the barrier constraint \eqref{eq:drone_cbf} which is adjusted to prevent the slung mass from contacting the ground and which has $\alpha = 0.995$.
For this test, because all controllers can only condition their disturbance on the current state (which does not include the position or velocity of the slung load), the disturbances appear to be random and high variance, since they are large and depend on the history of states visited previously. Here we compare only our proposed method \eqref{eq:orio} and the MLP; as expected, in this noisy case the our CVAE-based method performs remarkably better as seen in Fig. \ref{fig:hanging-mass}, and has no safety violations. A video of the experiments can be found with our code \href{\codebase}{here}. This experiment demonstrates both that the CVAE can learn an accurate stochastic model of highly noisy dynamics (including trajectories where the slung load reached nearly 90 degree angles dynamics), and also is necessary to ensure safety in such noisy cases.

\section{Conclusions}

We present a unified framework for risk-senstive control that combines CVAEs, which learn stochastic disturbance models from trajectory data, with DTCBFs, which provide probabilistic safety guarantees for stochastic systems. We demonstrate the real-time utility of this framework by running the full pipeline (after training) at 100Hz onboard a quadrotor drone performing aggressive flight, including a free fall and flight with a slung load.

Future work involves extending the method to handle other forms of uncertainty such as those generated by perception, reduced order models, contact dynamics, or human-robot-interaction. 



\section{Appendix} \label{appendix}
First we provide additional details for Theorem \ref{thm:drone_safety} where we consider some function $h$ as in \eqref{eq:drone_cbf}: 
\begin{align}
    h(\mb{x}) = (C - \bs{\zeta}^\top \mb{P} \bs{\zeta} )  - \lambda ( 1 - \mb{e}_z \mb{R}(\mb{q}) \mb{e}_z) \label{eq:drone_cbf_apdx}
\end{align}
where $C > 0$, $\lambda >0$, $\mb{x} = (\mb{p} \in \R^3, \mb{q} \in \textup{SO}(3), \mb{v} \in \R^3)$, the position is $\mb{p} = [x, y, z]$, the velocity is $\mb{v} = [v_x, v_y, v_z]$, $\bs\zeta = [z - z_0, v_z ]$ for some $z_0 \in \R$, $R(\mb{q})$ is the rotation matrix representation of the orientation $\mb{q}$,  and $\mb{e}_z = [0, 0, 1]^\top $. 

Additionally, we require that $\mb{P}\in \R^{2\times2}_{\succ 0}$ satisfies the Discrete-time Algebraic Ricatti Equation (DARE) for the discrete time double integrator system with time step $\Delta_t > 0 $: 
\begin{align}
    \displaystyle
    \bs{\zeta}_{k+1} = \underbrace{\lmat 1 & \Delta_t \\ 0 & 1 \rmat}_{\mb{A}} \bs{\zeta}_k  + \underbrace{\lmat 0 \\ \Delta_t \rmat}_{\mb{B}} \tilde{u}.  
\end{align}
That is, $\mb{P}$ satisfies the DARE equation \cite{ran1988existence}: 
\begin{align*}
    \mb{P} = \mb{A}^\top \mb{P} \mb{A} - (\mb{A}^\top \mb{P}\mb{B})(R + \mb{B}^\top \mb{P}\mb{B})^{-1}(\mb{B}^\top \mb{P} \mb{A}) + \mb{Q}      
\end{align*}
for some $R >0 $ and $\mb{Q}\in\R^{2\times 2}_{\succ 0} $. We note that using the input 
\begin{align}
    k_D(\bs{\zeta}) = -(\mb{B}^\top\mb{P}\mb{B} + R)^{-1}(\mb{B}^\top \mb{P} \mb{A})\bs{\zeta}
\end{align} 
ensures that the system satisfies: 
\begin{align}
    \bs{\zeta}_{k+1}^\top \mb{P}_{k+1} \bs{\zeta} \leq  \bs{\zeta}_k^\top (\mb{P} - \mb{Q}) \bs{\zeta}_k. 
\end{align}

For this proof we consider the Euler-approximated dynamics of $\mb{x}$ where the orientation is represented using a rotation matrix. In particular, the approximated dynamics are: 
\begin{align}
    \mb{p}_{k+1} & \approx \mb{p}_k + \Delta_t\mb{v}_k + \bs{\delta}_1\\
    \mb{R}(\mb{q}_{k+1}) & \approx \mb{R}(\mb{q}_k) + \Delta_t \mb{R}(\mb{q}_k) [\bs{\omega}_k]_\times + \bs{\delta}_2\\
    \mb{v}_{k+1} & \approx \mb{v}_k - \Delta_t \mb{e}_z g + \Delta_t  R(\mb{q}_k) \mb{e}_z\tau_k + \bs{\delta_3}
\end{align}
where $\mb{u}_k = (\tau_k \in \R, \bs{\omega}_k\in \R^3)$, $g\in \R$ is gravitational acceleration, and the disturbance is $\bs{\delta} = (\bs{\delta}_1 \in \R^3, \bs{\delta}_2 \in \mathfrak{so}(3), \bs{\delta}_3 \in \R^3)$. These are collected in the approximated dynamics $\mb{x}_{k+1} \approx \bs{F}_{\textup{eul}}(\mb{x}_k, \mb{u}_k) + \bs{\delta}$. The effect of using this approximation instead of the true dynamics is analyzed in \cite{tayal_control_2023}. 

Now we restate the theorem: 
\begin{theorem*}
    Consider $h$ as in \eqref{eq:drone_cbf_apdx} and some $\alpha \in [0,1)$ such that $(\mb{Q} - (1 - \alpha) \mb{P})\in \mathbb{S}^n_{\geq 0}$. If $C > 2 \lambda$, then there exists $\mb{u}  \in \R^4$ and $M_{\bs{\delta}}, M_{\bs{c}}>0$ such that 
    \begin{align}
        h(\mb{F}_{\textup{Eul}}(\mb{x}, \mb{u}) + \bs{\delta})  + c \geq \alpha h(\mb{x}) 
    \end{align} for all $  \bs{\delta} $ such that $\Vert \bs{\delta} \Vert \leq M_\delta,  c \leq M_c $ and all $\mb{x}$. 
\end{theorem*}

The proof of this theorem relies on the differential-flatness of the quadrotor system by considering two cases: one when the drone's thrust vector is at least partially aligned with the $z$-direction and the other when it is normal to the $z$-direction. By finding feasible control actions in both cases, we prove the feasibility of the constraint in general. 

\begin{proof}
    Let $\varphi(\mb{R}(\mb{q})) \triangleq \mb{e}_z^\top \mb{R}(\mb{q}) \mb{e}_z$.
    \begin{itemize}
        \item     If $\varphi(\mb{R}(\mb{q}))  \neq 0 $, then $\tau = \frac{ 1}{\varphi(\mb{R}(\mb{q})) } \left( g +  k_D(\bs{\zeta}) \right) $ with $[\bs{\omega}]_\times = - \frac{1}{\Delta_t}\mb{R}(\mb{q})^\top[\bs{\delta}_2]_{\times}$ yields: 
    \begin{align}
    \displaystyle
        & h(\mb{F}_{\textup{Eul}}(\mb{x}, \mb{u}) + \bs{\delta}) \\
        & = C - \underbrace{\left( \mb{A}\bs{\zeta} + \mb{B}k_D(\bs{\zeta}) + \lmat \bs{\delta}_1 \\ \bs{\delta}_3\rmat \right)}_{\triangleq \mb{F}_{\bs{\zeta}}(\bs{\zeta}, k_D(\bs{\zeta}), \bs{\delta})}^\top P \mb{F}_{\bs{\zeta}}(\bs{\zeta}, k_D(\bs{\zeta}), \bs{\delta}) \nonumber \\
        &  \quad \quad \quad - \lambda (1 - \mb{e}_z\mb{R}(\mb{q}) \mb{e}_z) \\ 
        & \geq h(\mb{x}) + \bs{\zeta}^\top Q \bs{\zeta}  - \gamma\left(\left\Vert \lmat \mb{e}_z^\top \bs{\delta_1} \\ \mb{e}_z^\top \bs{\delta_3} \rmat \right\Vert \right)   
    \end{align}
    The inequality and the existence of $\gamma\in \mathcal{K}$ is guaranteed through the use of the DARE controller \cite{jiang_input--state_2001}. Since $(\mb{Q} - (1 - \alpha) \mb{P}) \in \mathbb{S}_{\geq 0}^n$ and $\mb{R}(\mb{q}) \in \textup{SO}(3)$, we can expand this inequality as: 
    \begin{align}
    \displaystyle
        &=\alpha h(\mb{x}) + \bs{\zeta}^\top (\bs{Q}- (1 - \alpha) \mb{P})\bs{\zeta} \nonumber  \\
        & \quad  + (1- \alpha)(C - \lambda \varphi(\mb{R}(\mb{q}))) -  \gamma\left(\left\Vert \lmat \mb{e}_z^\top \bs{\delta_1} \\ \mb{e}_z^\top \bs{\delta_3} \rmat \right\Vert \right) \nonumber  \\  
        &\geq \alpha h(\mb{x}) + (1 - \alpha)( C - 2 \lambda ) -  \gamma\left(\left\Vert \lmat \mb{e}_z^\top \bs{\delta_1} \\ \mb{e}_z^\top \bs{\delta_3} \rmat \right\Vert \right)   \nonumber
    \end{align}
    Pick $0 < M_{\bs{\delta}} < \gamma^{-1}((1 - \alpha)( C - 2 \lambda )) $ and $0<M_c \leq (1 - \alpha)(C - 2 \lambda) - \gamma(M_\delta) $, then
    $h(\mb{F}_{\textup{Eul}}(\mb{x}, \mb{u}) + \bs{\delta})  \geq \alpha h(\mb{x}) + c$
    as desired. 
    
        \item     If $\varphi(\mb{R}(\mb{q})) = 0 $, then 
        \begin{align}
            & h(\mb{F}_{\textup{Eul}}(\mb{x}, \mb{u}) + \bs{\delta}) \nonumber \\
            & = C - \bs{\zeta}_{k+1}^\top \bs{P} \bs{\zeta}_{k+1} - \lambda ( 1 - \mb{e}_z (\mb{R}(\mb{q}) + \bs{\delta}_3)  \mb{e}_z) \nonumber \\
            & \quad \quad \; + \lambda \Delta_t \mb{e}_z \mb{R}(\mb{q})[\bs{\omega}]_\times\mb{e}_z.   
        \end{align}
        Thus, the effect of $\bs{\omega}$ on $h(\mb{F}_{\textup{Eul}}(\mb{x}, \mb{u}) + \bs{\delta}) $ is         
        \begin{align}
             \lambda \Delta_t \mb{e}_z \mb{R}(\mb{q}) [\bs{\omega}]_\times \mb{e}_z =  \lambda \Delta_t \mb{e}_z \mb{R}\bs{\varpi} 
        \end{align}        
    where $\bs{\varpi} = \lmat \omega_x, & \omega_y, & 0 \rmat^\top  $. Since $\varphi(\mb{R}(\mb{q})) = 0$,  $\mb{e}_z\mb{R}_k $ is perpendicular to $\mb{e}_z$ and thus there exists $\omega_x, \omega_y$ such that: 
        $ \lambda \Delta_t \mb{e}_z \mb{R}(\mb{q})[\bs{\omega}]_\times \mb{e}_z = \psi $ for any $\psi \in \R $.

    Now choose $\tau = 0 $ and $\bs{\omega} $ such that: 
    \begin{align}
    \displaystyle
        & h(\mb{F}_{\textup{Eul}}(\mb{x}, \mb{u}) + \bs{\delta}) \nonumber \\
        & = C - \lambda ( 1 - \mb{e}_z (\mb{R}(\mb{q}) + \bs{\delta}_3)  \mb{e}_z) + \psi \nonumber \\
        & \quad \quad - \underbrace{\left(\mb{A}\bs{\zeta} + \lmat \bs{\delta}_1 \\ \bs{\delta}_2 \rmat \right)^\top \mb{P} \left(\mb{A}\bs{\zeta} + \lmat \bs{\delta}_1 \\ \bs{\delta}_2 \rmat \right)}_{\triangleq p(\bs{\delta})}\\ 
        & \geq C - \lambda (2 + M_\delta ) - \sup_{\Vert\bs{\delta} \Vert \leq M_\delta } p(\bs{\delta}) + \psi
    \end{align}
    Since $\mb{P}$ is positive definite,  $\sup_{\Vert\bs{\delta} \Vert \leq M_\delta } p(\bs{\delta})$ is bounded. Furthermore since we can choose $\bs{\omega}$ such that $\lambda \Delta_t \mb{e}_z \mb{R}(\mb{q})[\bs{\omega}]_\times \mb{e}_z = \psi$ for any $\psi$, choose 
    \begin{align}
    \displaystyle
         \psi = -\left( C - \lambda (2 + M_\delta ) - \sup_{\Vert\bs{\delta} \Vert \leq M_\delta } p(\bs{\delta}) \right) + \alpha h(\mb{x}) + c \nonumber
    \end{align}
    then we get $h(\mb{F}_{\textup{Eul}}(\mb{x}, \mb{u}) + \bs{\delta})  \geq \alpha h(\mb{x}) + c$ as desired. 
    \end{itemize}
    Since there exists a controller that satisfies the constraint for all $\mb{x} $ such that $\varphi(\mb{R}(\mb{q})) = 0$ and all $\mb{x}$ such that  $\varphi(\mb{R}(\mb{q})) \neq 0 $, the constraint is feasible for all $\mb{x}$.  
\end{proof}

\bibliographystyle{IEEEtran}
\bibliography{references}

\end{document}